\documentstyle[12pt]{article}
\begin{document}
\title{Many-Electron Systems with Constrained Current}
%\title{ Inhomogeneous interacting electron gas with constant  current}
%\title{Variational Formulation of Constant Current Quantum Mechanics}
\author{D.S. Kosov$^{1,2}$ and J.C. Greer$^1$}
\date{}
\maketitle
\begin{center}
$^1$ NMRC, University College,
Lee Maltings, Prospect Row, Cork, Ireland \\
\vspace{0.1cm}
$^2$ Institute of Physical and Theoretical Chemistry, 
J.W. Goethe University, Marie-Curie-Str.11, D-60439, Frankfurt,
Germany
\end{center}

\begin{abstract}
A formulation for transport in an inhomogeneous, interacting
electron gas is described. Electronic current is induced by 
a constraint condition imposed as a vector Lagrange multiplier. 
Constrained minimization of the total energy functional on the 
manifold of an arbitrary constant current leads to a many-electron 
Schr\"odinger equation with a complex, momentum-dependent potential. 
Constant current Hartree-Fock and Kohn-Sham approximations are 
formulated within the method and application to 
transport for quantum wires is developed. No appeal is made to 
near equilibrium conditions or other approximations allowing development
of a general ab initio electronic transport formulation. 
\end{abstract}

In recent years, there has been considerable advances towards 
the development of accurate theoretical models to deal with 
electron transport through atomic and molecular 
wires \cite{nitzan01}. 
This activity has been 
largely spurred on by development of experimental techniques 
to form atomic scale electrical contacts such as scanning tunneling 
microscopy \cite{mirkin92} and mechanically controllable break 
junctions \cite{ruiten96}. Measurement
of current-voltage characteristics have been recently performed on
a single benzene-1,4-dithiolate molecule \cite{reed97}, 
buckminsterfullerene \cite{joachim97}, 
individual atoms \cite{scheer98}, and DNA 
strands \cite{porath2000}. The mechanisms for charge transport through 
these low-dimensional structures remain largely unexplored and 
unexplained. Several approaches have been 
developed on the basis of the extended H\"uckel \cite{mujica94} and 
density functional theory within a Lippmann-Schwinger 
formalism \cite{diventra2000} or nonequilibrium Green's functions \cite{taylor2001}. 
In these approaches, the system is partitioned into three parts: 
molecular device (or wire) and two electron reservoirs. 

The interaction between contacts and wire represents the kernel for the 
Lippmann-Schwinger equation, and serves generally
to define the non-interacting system for Green's function approaches. 
Recent studies indicate the exact nature of the molecule-contact bonding are 
critical in predicting the correct order of magnitude for the currents with 
applied bias revealing that accurate electronic structure methods must be
incorporated into quantum transport methods for accurate predictions and
analysis of current-voltage relationships.  

In this letter, we present a variational approach to charge transport in correlated 
electron systems. We begin by introducing a many-electron Schr\"odinger equation
which provides exact wavefunctions on the manifold of a given current distribution. 
The many-electron Schr\"odinger equation is developed without any restrictions to the 
treatment of electron-electron correlations, whether as injected current or part of the
molecular device. We next present specific approximations corresponding to the Hartree-Fock and
configuration interaction treatments of the many-electron problem.
Next, the Kohn-Sham approximation to density functional theory is extended to 
the inhomogeneous electron gas with a fixed current at zero external magnetic field. 

We begin by developing a general many-body theory for an inhomogeneous electron gas
with arbitrary and {\it static} current density. 
For an inhomogeneous, interacting $N$-electron gas 
in an external potential $v({\bf r})$, the
ground state energy is given by the following energy functional 
(atomic units are  employed throughout the paper unless
otherwise mentioned):
\begin{equation}
\label{Energy}
E[\rho, \Gamma] = \langle \Psi | H|\Psi \rangle =
\int h_0({\bf r}) \rho({\bf r},{\bf r}')|_{{\bf r}={\bf r}'}\, d{\bf r}+
\frac{1}{2}\int\frac{1}{|{\bf r}_1 -{\bf r}_2|} \Gamma({\bf r}_1,{\bf r}_2;{\bf r}_1,{\bf r}_2) 
d{\bf r}_1 d{\bf r}_2
\end{equation}
where $h_0({\bf r}_i)$ is the one-body electron operator which includes the electronic 
kinetic energy 
and the one-body electron interactions with external potentials. 
 We have introduced the one- and two-electron 
reduced density matrices: 
\begin{eqnarray}
\rho({\bf r};{\bf r}') &=& N \int \Psi   ({\bf r},{\bf r}_2,\ldots,  {\bf r}_N)  
                                  \Psi^* ({\bf r}',{\bf r}_2,\ldots, {\bf r}_N) 
\prod_{i=2}^N d{\bf r}_i \\
\Gamma({\bf r},{\bf r}';{\bf r},{\bf r}') &=& N(N-1) \int 
\Psi   ({\bf r},{\bf r}',\ldots,  {\bf r}_N)  
\Psi^* ({\bf r},{\bf r}',\ldots, {\bf r}_N) \prod_{i=3}^N d{\bf r}_i , 
\end{eqnarray}
where $\Psi$ is the many-electron wave-function

The fundamental quantity in our approach is the electron current density
\begin{eqnarray}
{\bf j}({\bf r})&=& \frac{1}{2i} 
\left[ \nabla_{{\bf r}}-\nabla_{{\bf r}'}\right]\rho({\bf r},{\bf r}')|_{{\bf r}={\bf r}'} .
\label{j1}
\end{eqnarray}
The current density is constrained to have specified values at prescribed points in 
position space, or more generally a functional $G[{\bf j}]({\bf r})$ is required to 
be zero at defined positions ${\bf r}$.
The task is to minimize the energy functional $E[\rho,\Gamma]$ subject to the constraint
$G[{\bf j}]({\bf r})=0$.
Altyhough the constraint formulated in this manner is nonholonomic, it may be included 
into a variational functional via pointwise vector Lagrange multipliers 
${\bf a}({\bf r})$. 
The constraint is explicitly 
achieved by introduction of an auxiliary functional
\begin{equation}
\Omega[ \rho, \Gamma, {\bf j}, {\bf a}]= E[\rho,\Gamma]-
\mu \left( \int d{\bf r } \rho({\bf r}) 
-N\right) + \Lambda[{\bf j, {\bf a}}] 
\label{omega}
\end{equation}
The first two terms are standard in the variational formulation of quantum many-body theory
with the first giving the total energy and the second is introduced to constrain 
particle number
(or likewise, to introduce the orthonormality constraint for the wavefunction).
The third term has been introduced to impose the current constraint.
The functional $\Lambda[ {\bf j},{\bf a}]$ depends linearly 
on ${\mit \bf {\bf a}} ({\bf r}) $ and 
is designed to insure the following condition is maintained: 
\begin{equation}
\frac{\delta \Lambda[{\bf j}, {\bf a}]}{\delta {\bf a}({\bf r})}= 
G[{\bf j} ]( {\bf r}).
\end{equation}
Variation of $\Omega[ \rho, \Gamma, {\bf j}, {\bf a}]$ with respect to the Lagrange multiplier 
${\bf a} $ provides an additional constraint equation. 
One can view the ${\bf a}({\bf r})$ 
as an infinite set of Lagrange multipliers, each associated with $ G[{\bf j}]({\bf r})=0$
at a point ${\bf r}$. The variation of $\Lambda[{\bf j}, {\bf a}]$ 
over the current density 
${\bf j} ({\bf r}) $ yields a vector field:
\begin{equation}
{\bf A}({\bf r})=\frac{\delta \Lambda[{\bf j}, {\bf a}]}{ \delta {\bf j} ({\bf r})}.
\end{equation} 
Extrema of the auxiliary functional $\Omega[ \rho, \Gamma, {\bf j}, {\bf a}]$ corresponds to the following many-electron
Schr\"odinger 
equation
\begin{equation}
\label{ConstrainedSWE}
\left(
\sum_j \left\{ h_0({\bf r}_j) + 
\frac{1}{2i}\left[ \nabla_j, {\bf A}({\bf r}_j) \right ]_+ \right\} +
\sum_{i<j} \frac{1}{|{\bf r}_i -{\bf r}_j|} 
\right) \Psi =E\, \Psi
\end{equation}
where we have defined the energy $E=N\mu$. The anti-commutator term,
\begin{equation}
\left[ \nabla, {\bf A}({\bf r}) \right ]_+ = \nabla  {\bf A}({\bf r}) 
                                        + 2{\bf A}({\bf r})  \nabla,
\end{equation}
gives rise to an additional single-electron imaginary potential arising directly 
from the constraint on the 
current density. This additional constraint potential forces the 
many-electron wave function $\Psi $ 
to be irremovably complex and enforces  
the required current density distribution at extrema. 

A physical interpretation of the vector field $ {\bf A }({\bf r})$ follows if 
we re-write the eq.(\ref{ConstrainedSWE}) in the following form
\begin{equation}
\left(
\sum_j \left\{ \frac{1}{2 } \left(-i \nabla_j + {\bf A}({\bf r}_j) \right)^2 
-\frac{1}{2}  {\bf A}({\bf r}_j) {\bf A}({\bf r}_j)+  v({\bf r}_j) \right\} +
\sum_{i>j} \frac{1}{|{\bf r}_i -{\bf r}_j|} 
\right) \Psi =E \Psi .
\label{schr2}
\end{equation}
The current constraint has introduced terms equivalent to a
vector potential of an external magnetic field \cite{vignale87}, however, the vector field 
is a functional of the $\Psi $ and has to be dtermined self-consistently 
from the Eq.{\ref{schr2}} in our case.

The constrained many-electron Schr\"odinger equation is not yet in a form 
allowing for a solution to be found as the 
the Lagrange multiplier ${\bf a}({\bf r})$ and the vector field ${\bf A}({\bf r})$,
which is a functional 
of ${\bf a}({\bf r})$, for a specified current are not given. 
An expression for ${\bf a}({\bf r})$ can not be found 
until an explicit form of constraint functional $\Lambda[{\bf j}, {\bf a}]$ is first introduced.
The motivation for the present study is to desire to develop the basic many-body formalism 
for the calculation of a direct current $I$ through a quantum wire. To achieve this aim, 
we specify the constraint on the current density distribution in the following form:
\begin{equation}
\int dy\, dz\, j_x({\bf r})=I_x
\label{constraint1}
\end{equation}
Within this description, net current flow is aligned along
the $x-$axis. 
This is a simple geometric arrangement to specify current flow for quantum wire
and can be easily extended
to more complex topologies.
For the case specified, the functional $ \Lambda[{\bf j}, {\bf a}] $ 
takes the following form:
\begin{equation}
\Lambda[ {\bf j},{\bf a}] = 
\int  dx {a}_x(x)( \int dy dz\, j_x({\bf r})-I_x)
\end{equation}
The vector field induced by the presence of the constraint on the current density 
has spatial dependence and nonzero projection only along the $x$-axis:
\begin{equation}
{\bf A }({\bf r}) = 
\frac{\delta \Lambda[ {\bf j},{\bf a}]}{\delta {\bf j}({\bf r})} 
= ({a}_x(x), 0,0)
\end{equation}
Therefore, for a quantum wire with a direct current the vector field ${\bf A}({\bf r})$ 
coincides with 
the vector Lagrange multiplier ${\bf a}({\bf r}) $. Generally, if the constraint functional 
$\Lambda[ {\bf j},{\bf a}]$
is a linear functional of ${\bf j}({\bf r})$, then the vector field ${\bf A}({\bf r})$ and the 
Lagrange multiplier at each point ${\bf r}$ are equivalent.

We next 
solve for the Lagrange multiplier
${a}_x(x)$
and obtain
\begin{equation}
{a}_x(x) = \frac{1}{2I_x} 
\left\{ \int dy dz \int \prod_{i=2}^N d{\bf r}_i 
\left[ \Psi^* H \Psi +\Psi H \Psi^* \right]
- 2 \mu \rho_x(x) 
\right\},
\end{equation}
where we have introduced the quantity
\begin{equation}
\rho_x(x) = \int dy dz \rho ({\bf r}).
\end{equation}

Suppose now that the many-electron wave-function
$\Psi$ is approximated as Slater determinant of N orthonormal 
single-electron orbitals $\psi_i({\bf r})$. 
The use of the Slater determinant enables us to write the electron current density 
(\ref{j1}) as the 
sum of electron current through all occupied orbitals
\begin{equation}
{\bf j}({\bf r})=\frac{1}{2i} \sum_{i}^{occ} \left( \psi_i^* ({\bf r}) \nabla \psi_i ({\bf r}) 
- \psi_i ({\bf r}) \nabla \psi_i^*({\bf r}) \right)
\end{equation}
Likewise, as for the derivation of the exact many-electron Schr\"odinger equation, the 
current distribution is
fixed via $G[{\bf j}]({\bf r})=0 $. The constraint is included into 
Hartree-Fock minimization via an additional Lagrange multiplier 
$\Lambda[{\bf j},{\bf a}]=\Lambda[\psi, \psi^*,{\bf a}]$:
\begin{equation}
\Omega[ \psi, \psi^*,{\bf a}]= E[\psi, \psi^*]  + 
\Lambda[ \psi, \psi^*,{\bf a}] 
\label{omegaHF}
\end{equation}
Minimization of eq.(\ref{omegaHF}) subject to the orthogonalization condition
$ \int\, d{\bf r} \psi_i^*({\bf r})\psi_j({\bf r}) =\delta_{ij} $ results in a 
Hartree-Fock equations with fixed current density distribution
\begin{equation}
\label{ConstrainedHF}
\left( \hat{H}_{HF} + \frac{1}{2i}\left[ \nabla, {\bf A}({\bf r}) \right ]_+  
\right) \psi_i({\bf r}) =E_i\, \psi_i({\bf r}) \; ,
\end{equation}
where $\hat{H}_{HF}$  is the Hartree-Fock operator which contains the electron kinetic energy,
external potential, 
Hartree potential and Fock exchange operator, ${\bf A}$ is again an additional vector 
field induced by the current constraint. 
The configuration interaction (CI) representation follows directly
if we substitute for the many electron wave function $\Psi$ a linear superposition of Slater 
determinants or spin coupled determinants. The CI expansion is substituted into eq. \ref{ConstrainedSWE}
and variation with respect to the expansion coefficients  
yields a modified matrix eigenvalue problem, whereby
the interaction matrix elements are supplanted with the additional one-body potential terms arising
from the constraint potential $\frac{1}{2i}\left[ \nabla, {\bf A}({\bf r}) \right ]_+$.

We next turn to the task of incorporating the current constraint into the 
density functional theory (DFT) \cite{hohenberg64}. 
We describe an extension of the Kohn-Sham 
formulation of DFT \cite{kohn65} to systems with a non-zero current density
distribution.
For an inhomogeneous, interacting electron gas in external potential $v({\bf r})$, the
ground state energy is given by the Hohenberg-Kohn energy functional \cite{hohenberg64}:
\begin{equation}
E_{HK}[\rho]=T_o [\rho ]+ \int d{\bf r} v({\bf r}) \rho({\bf r})+
\frac{1}{2} \int \int \frac{ \rho({\bf r}) \rho({\bf r}')}
{|{\bf r}- {\bf r}'|} +E_{xc}[\rho ]
\end{equation}
The first term, $T_0$, is the kinetic energy functional of $N$ noninteracting electrons with 
given density and current density distribution. The second and third terms, 
are the interaction energy with 
external potential and the electrostatic interactions. 
The last term, $E_{xc}[\rho ]$, is the exchange and correlation energy functional.

We follow closely to the derivation of the many-electron Scr\"odinger equation 
described in the beginning of this letter.
The distribution for the current density vector is fixed through a
functional $G[{\bf j}]({\bf r})=0$ at certain points ${\bf r}$.
The constraint is included into DFT variational determination of the
charge density via a Lagrange multiplier ${\bf a} ({\bf r})$:
\begin{equation}
\Omega[ \rho, {\bf j}, {\bf {\bf a}}]= E_{HK}[\rho] - \mu \left( \int d{\bf r } \rho({\bf r}) -N\right) + 
\Lambda[ {\bf j},{\bf a}] 
\label{omegaHK}
\end{equation}

Following the Kohn and Sham approach \cite{kohn65}, we introduce a reference 
fermion system with orthonormal single-particle orbitals $\psi_i({\bf r})$ and 
occupition numbers $n_i$ to reproduce 
the charge and current densities:
\begin{equation}
\rho( {\bf r})= \sum_i n_i \psi_i^*({\bf r}) \psi_i({\bf r})
\nonumber 
\end{equation}
\begin{equation}
{\bf j}({\bf r})=\frac{1}{2i} \sum_{i} n_i \left( \psi_i^* ({\bf r}) \nabla \psi_i ({\bf r}) 
- \psi_i ({\bf r}) \nabla \psi_i^*({\bf r}) \right)
\end{equation}

In analogy to the Kohn-Sham approach, we consider variation of the auxiliary functional 
$\Omega[ \rho, {\bf j}, {\bf {\bf a}}]$ eq. \ref{omegaHK} with respect 
to the single-electron orbitals $\psi_i({\bf r})$ to obtain the following 
single-electron self-consistent equations:
\begin{equation}
\left[
\frac{1}{2} \left(- i \nabla +  {\bf A}({\bf r})
\right)^2 
- \frac{1}{2} {\bf A}({\bf r})  {\bf A}({\bf r})+
  v({\bf r}) + \int d{\bf r} \frac{\rho({\bf r}')}{|{\bf r}-{\bf r}'|} 
+ \mu_{xc}[\rho]({\bf r})  \right] \psi_i({\bf r})= 
E_i \psi_i ({\bf r}),
\label{ks}
\end{equation}
the desired result.

In order to get an indication of how our approach works,
we solve the Kohn-Sham equation with fixed current (\ref{ks}) for the case
of a uniform electron gas. Consider an electron gas in a cubical box of volume $V =L^3$,
throughout which a background positive charge is uniformly spread out
rendering the system neutral (jellium model).
A direct current is applied along the x-axis
${\bf I}=(I_x,0,0)$.
The classical Coulomb electron-electron interaction,
the Hartree term, is completely
compensated by the electron interaction with positive charge density and by
the electrostatic energy of the positive background.
The vector field is constant for the translationaly invariant
uniform electron gas yielding the following
Kohn-Sham equation for the model:
\begin{equation}
\left( -\frac{1}{2} \Delta +\mu_{ex} + \frac{1}{i} {\bf A}\nabla
\right) \psi_k({\bf r}) =E_k\, \psi_k({\bf r})
\label{modks}
\end{equation}
with specification of the constraint taken in the form given by eq.(\ref{constraint1}).
A dispersion relation which is coupled to the constraint equation for the current density
may be obtained by inserting a
plane-wave solution $\psi_k ({\bf r}) =\frac{1}{\sqrt{V}} \exp(i {\bf k}{\bf r})$
into eq2.(\ref{modks}, \ref{constraint1}), with the result
\begin{equation}
E_{\bf k} = \frac{ k^2}{2 } +\mu_{xc} + A_x k_x
\label{ek}
\end{equation}
\begin{equation}
\frac{1}{L} \sum_{k_x}^{occ} k_x =I_x
\label{I}
\end{equation}
From the system of equations eqs.(\ref{ek}-\ref{I})
the exact
expression for the vector field $A$ can be deduced:
\begin{equation}
A = - \frac{I_x}{\rho} \;,
\end{equation}
with electron density $\rho= N/L$.
Given the vector field, we can write
the current density dependent dispersion relation of the uniform
electron gas with applied direct current $I_x$:
\begin{equation}
E_{\bf k} = \frac{k^2}{2} +\mu_{xc} - \frac{I_x}{\rho}  k_x
\end{equation}
The presence of the constrained current lifts the degeneracy of the $k$ and $-k$ states
and results in the energy gap between  the electron moving in the direction of net current flow, 
i.e. with positive $k_x$, and electrons moving in
the opposite direction.

We finally demonstrate application of the method with a simple numerical example. We consider
a one-dimensional system with current injection into a fixed, external potential. 
Trial states are taken to be pure plane wave states and these states are then allowed
to relax to account for the presence of an external potential, taken to be a simple
Gaussian form. 
In fig. 1, the solutions to the Schr\"odinger equation are shown with the
current fixed by the constraint potential, i.e. no boundary conditions are imposed
other than through the constraint condition. Within the figure, the changes in real
and imaginary components of the wave function relative to a pure plane wave state are plotted.
The constraint re-arranges the solution to the Schr\"odinger equation allowing for the changes
due to the current encountering the external potential, while maintaining a constant current
fixed at the value specified by the constraint condition.

In this letter, we have given a variational formulation of quantum electronic
transport. There is no resort to any 
imposed conditions other than to constrain the current distribution. Thus the formulation
is equally valid for highly correlated systems and for all nonequilibrium current regimes.
The formulation has been specified for general many-body theory and cast into a form
suitable for the Hartree-Fock and configuration interaction methods. We next showed how
to introduce the constraints into density functional theory. The method as such is quite 
general and applicable to all common approaches to electronic structure theory.

For the case of one-dimensional transport in a two-terminal system, we have introduced a
physical constraint which allows for an explicit determination of the Lagrange multipliers
required to fix the current flow. With the Lagrange multipliers in hand, it is possible
to determine the complex potential needed for solution of the many-body wavefunction, or 
electronic density, 
satisfying the physical boundary conditions introduced by a constant current flow in and out 
of a region. The formulation of the problem has been chosen for greatest compatibility with the
methods in common use for electronic structure theory determination, and results in a completely
variational formulation of the quantum electronic transport problem.
\\
\\
\noindent
{\bf Acknowledgment} This work has been funded by the European Union through the Information
Society's Technology (IST) programme, within the Future and Emerging Technologies Advance Research
Initiative's NANOTCAD project (IST-1999-10828).
 
\clearpage
\noindent
{\bf Figure caption}
\\
\noindent
figure 1- Wave function shifts from an attractive Gaussian potential 
$ V_{ext} = V_0 \exp (-(x-\mu)^2/\sigma^2) $ are shown as the difference in
the wavefunction $\Psi = \phi + i \xi $ relative to a plane wave state $\Psi_0$ in box normalized
units. All other quantities expressed in atomic units. Incident energy
is E = 2.0 a.u., well depth V$_0$=-0.1 a.u., well 
breadth $\sigma=0.04\pi$.
Shift in the real component $\phi-\phi_0$ are displayed with the heavy lines, shift in 
the imaginary component $\xi -\xi_0$ are displayed with the lighter lines.

\clearpage
%%%%%%%%%%%%%%%%%%%%%%%%%%%%%%%%%%%%%%%%%%%%%%%%%%%%%%%%%%%%%%%%%%%%%%%%%%%%%

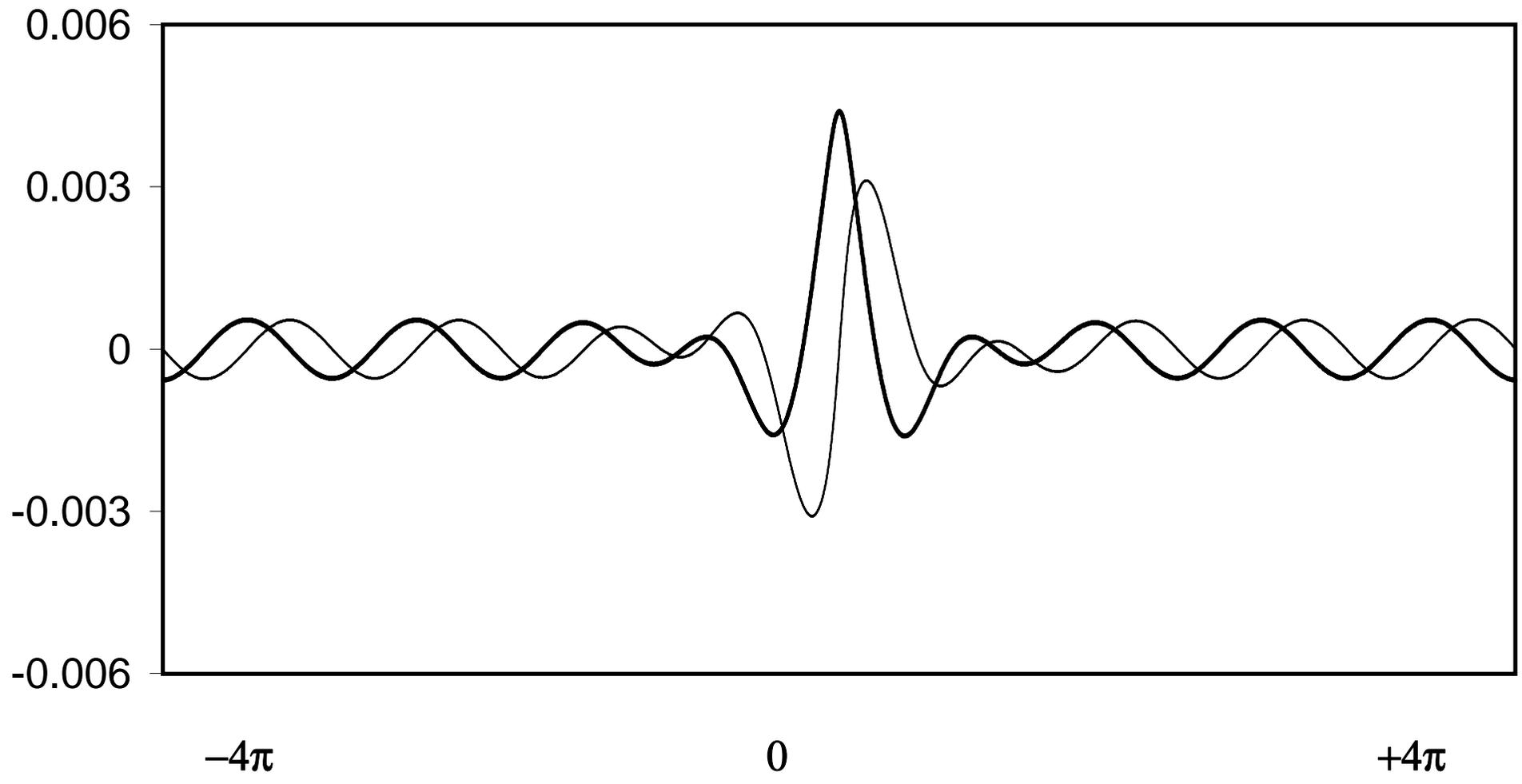
\end{document}